# Automatic Summarization of Soccer Highlights Using Audio-visual Descriptors

A. Raventós,
Signal Theory and Communications Department
UPC-BARCELONATECH
Esteve Terradas, 7
08860 Castelldefels, Spain
<arnau.raventos@gmail.com>

R. Quijada,
Signal Theory and Communications Department
UPC-BARCELONATECH
Esteve Terradas, 7
08860 Castelldefels, Spain
<raulq.f@gmail.com>

Luis Torres
Signal Theory and Communications Department
UPC-BARCELONATECH
Jordi Girona 1-3
08034 Barcelona, Spain
<luis.torres@upc.edu>

Francesc Tarrés
Signal Theory and Communications Department
UPC-BARCELONATECH
Esteve Terradas, 7
08860 Castelldefels, Spain
<francesc.tarres@upc.edu>

Corresponding author
Luis Torres
<luis.torres@upc.edu>
Phone: +34 93 401 64 49
Fax: +34 93 401 72 00

# Automatic Summarization of Soccer Highlights Using Audio-visual Descriptors

A. Raventós, R. Quijada, L. Torres, F. Tarrés
Signal Theory and Communications Department
UPC-BARCELONATECH

**Abstract[1]**

*Automatic summarization generation of sports video content has been object of great interest for many years. Although semantic descriptions techniques have been proposed, many of the approaches still rely on low-level video descriptors that render quite limited results due to the complexity of the problem and to the low capability of the descriptors to represent semantic content. In this paper, a new approach for automatic highlights summarization generation of soccer videos using audio-visual descriptors is presented. The approach is based on the segmentation of the video sequence into shots that will be further analyzed to determine its relevance and interest. Of special interest in the approach is the use of the audio information that provides additional robustness to the overall performance of the summarization system. For every video shot a set of low and mid level audio-visual descriptors are computed and lately adequately combined in order to obtain different relevance measures based on empirical knowledge rules. The final summary is generated by selecting those shots with highest interest according to the specifications of the user and the results of relevance measures. A variety of results are presented with real soccer video sequences that prove the validity of the approach.*

## 1. INTRODUCTION

The significant advances in image and video compression during the last decade have propelled the creation of vast amounts of visual content [1]. This tremendous explosion in visual content and the myriad of applications that have arisen, has spurred the need to develop advanced tools for content-based image and video indexing and retrieval [2], [3], [4], [5]. Among them, summarization of video sequences is considered of paramount importance in many applications. Some reviews and latest developments in the field can be found in [2], [6], [7], [8], [9].

In the particular but very important application of video sport sequences, summarization plays a key role. Usually, the sequences are analyzed and summarized manually which implies a laborious and exhausting task. Due to the great amount of available sequences and the amount of time required in this process, there is a need to provide, in particular, with automatic sports video sequences highlights summarization approaches. Due to its extraordinary popularity in many countries, soccer is one of the most important areas where sports video sequences summarization is being applied. A variety of approaches have been presented in the literature.

In addition, it has to be taken into account that for many years, low-level descriptors have been the only approach available for sports video sequences summarization and in general for video analysis. These low-level descriptors include, among others, statistical moments, shape, color, texture, and motion. However, it is well recognized that such information is not enough for uniquely discriminating across different visual content. Thus, the use of advanced information is required in order to obtain meaningful results. In particular, spatio-temporal analysis, video structure and syntax and video sequence events have gained a lot of attention [10], [11], [12], [13], [14].

---

[1] This research has been partially funded by the Spanish National Science Foundation (TEC2011-22989).

A very short review on soccer video summarization is introduced next. Zawbaa et al. present in [15] a system that firstly segments the whole video sequence into small video shots, then it classifies the resulted shots into different shot-type classes. Afterwards, the system applies a support vector machine and an artificial neural network algorithm for emphasizing important segments with logo appearance. Subsequently, the system detects vertical goal posts and the goal net. Finally, the most important events during the match are highlighted in a soccer video summary. Precision and recall rates are found over 90%. Fendri et all propose in [16] an approach that builds upon segmentation and indexation that rely on both low-level and text-based treatment of the soccer video. Results present precision and recall rates are found over 60%. Ekin et all present in [17] a propose a fully automatic and computationally efficient framework for analysis and summarization of soccer videos using cinematic and object-based features. The proposed algorithm runs in real-time, and, on the average, achieves a 90.0% recall and 45.8% precision rates, which are quite satisfactory for real time events. Lofti et all present in [18] an approach that rejects the shots containing non-significant events to summarize the video. The algorithm is adequate for real time soccer summarization and presents 100% of success rates for detecting specific events such as goals and penalties. Tabii et all introduce in [19] a new approach for the automatic extraction of summaries in soccer video based on shot detection, shot classification and a Finite State Machine technique. No recall either precision rates are presented, but numerical results are satisfactory.

It is in this context that a new approach for soccer video sequences highlights summarization is presented. The approach is based on sport video edition human-expertise used in commercial television. The system presented here defines the shot as the minimum unit for building up the summary. A video segmentation approach is introduced that relies on the detection of the shot transitions. Then, a score is assigned to every shot and those shots with the highest score are selected in order to build up the final summary. To qualify for scoring, shots are first passed through an analysis bank that computes several low-level and mid-level audiovisual descriptors. These descriptors define the video contents for every shot. Once the shot descriptors have been computed, they are passed through a multiple filtering process that will attempt to associate semantic meaning to each shot and will provide an overall score. The scheme is designed such as the user is able to specify, up to some degree, the type of contents appearing in the summary and its approximate duration.

This paper is divided in the following sections. Section 2 presents an overview of the overall system. The details of the algorithms used in each module are explained in Section 3. Section 4 presents the summary generator architecture and the analysis bank for the soccer highlights selection. Section 5 presents the results and performance of the system for the soccer scenarios, and finally, Section 6 contains the conclusions and future work.

## 2. OVERVIEW OF THE SYSTEM

The design of the automatic soccer game highlights summary generator presented in this paper is based on sport video edition human-expertise used in commercial television. TV sport summaries are made-up of a sequence of short shots that collect the essential events of interest, usually in a linear timely basis where different shots are presented in the same order that have occurred. It is also assumed that the summary is generated from the video feed that was produced during the live broadcasting of the game. No auxiliary cameras, player-follow shots or alternative views are supposed to be available to generate the highlights video summary.

Keeping this TV production style in mind, the minimum unit for building up the highlights video summary will be the video-shot defined as a series of continuous frames captured by a single camera that runs for a period of time. The final summary will be a concatenation of selected video-shots found in the original sequence. The overall diagram of the automatic highlights generator proposed in this paper is represented in Fig. 1. The objective is to assign a score to every shot and then select those shots with the highest score in order to build up the final summary. To qualify for scoring, shots are passed through an analysis bank that computes several low-level and mid-level audiovisual descriptors. Once the shot descriptors have been computed, the resulting xml files are finally passed to the highlights generation stage which, through a multiple filtering process, will attempt to associate semantic meaning to each shot and will provide an overall score. The final score of a shot may take into account not only its own descriptors but also the ones of its neighbor's shots. The user may interact with the system specifying the total duration and the percentage of every semantic filter in the final summary.

One of the key components in the architecture of the system is the video-shot segmentation module that produces an output xml file indicating the initial and ending time codes for every shot. The output of the video-shot segmentation stage is then used by the analysis bank to process the audio and video tracks of every shot and determines a set of audio and video descriptors that will be annotated in xml format and transferred to the highlights generation module which will generate the final highlights video summary. In addition, another key component of the summarization system is the use of the audio information that, combined with the video information, provides more robust and accurate results than using only the video information. In particular, a new algorithm to extract the whistle information found in the audio track is presented in Section 3.

The reliability of the video-shot segmentation is based on the detection of the transitions. In live sport broadcasting, transitions are mainly hard-cuts and cross dissolve. Hard-cuts are used during the action of the game while the latter may be used before or after the game or during the half time. The algorithm for the transition detection is explained in Section III.A and combines two off-the-shelf methods in order to achieve a good trade-off between complexity and performance with these two types of transitions, selecting one or other method depending on the part of the game video sequence that is being processed.

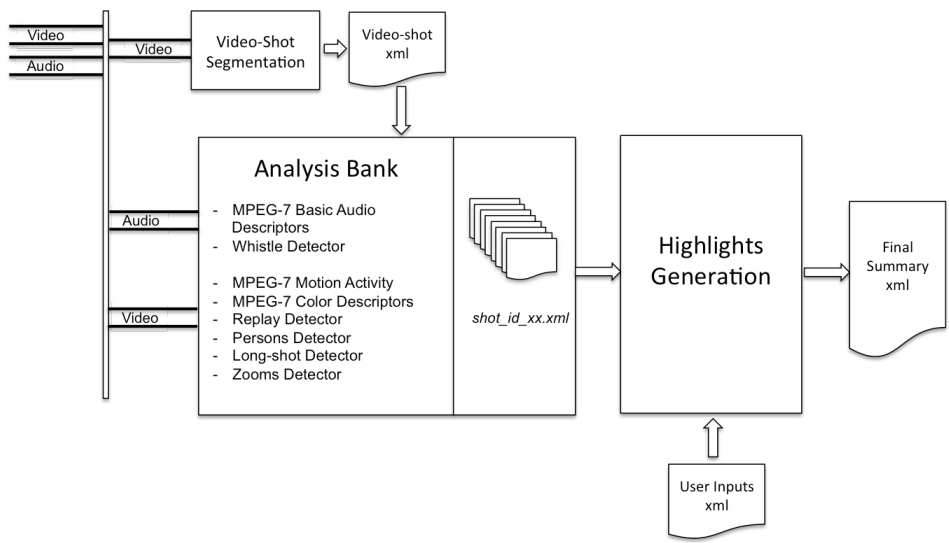

**Fig. 1** General diagram of the automatic highlights summary generator

The analysis bank includes a variety of analysis tools to extract the audio and video descriptors. Each of these tools produce an elementary xml file that describes the contents of the audio or and video sources with a single descriptor type. Some of these tools are applied directly to the segmented video or audio streams while others will only process a set of selected key-frames of the video sequence. In the latter case, the analysis tools are used at the output of a key-frame detector that selects a few frames (in most cases a single frame) that represent the shot. An example of a particular analysis tool is the person's detector.

The information collected from every shot include:

- Low-level audio MPEG-7 descriptors for every audio frame.
- The motion descriptors (Motion Activity and Camera Motion) associated to the shot.
- Detection of Zoom-in and Zoom-outs in the analyzed video shot.

- A list of key-frames that represent the video shot.
- Color Layout, Color Structure and Dominant Color MPEG-7 descriptors associated to every key-frame in the shot.
- Faces detected in the key-frames. Percentage of skin pixels in a key-frame.
- Long-shots detected in the key-frames.
- Replays detected in the broadcasted video feed time codes.
- Referee's whistle time codes.

The details of the algorithms used in each module are explained in Section 3.

The final stage of the system is the highlights summary generator module, that is composed of a filter bank that scores each shot with a set of elementary and advanced filters and the shot selection stage that generates the final summary by assessing the previous scores and a set of user parameters. The elementary filters find events of interest in the video shot. For example, one elementary filter can be the audio power increase inside the shot or the detection of the referee's whistle. Then, elementary filters are combined together in order to create a set of advanced filters that will give a final score to every shot. The objective is to assign positive or negative weights to each elementary filter in order to obtain the final score assigned to every shot. These filters are flexible enough to let the user interact with the system, specifying the relative importance of each elementary filter or, alternatively, to be statistically trained by a linear classifier. The user may define the number of advanced filters, their structure, the weighting of elementary filters, the percentage of every advanced filter included in the summary and the summary duration. The highlights summary generator module is presented in detail in Section 4.

In order to evaluate the performance of all algorithms of the proposed approach and the overall automatic highlights summary generator, three different video groundtruths have been generated. Each groundtruth contains the following soccer matches:

- Groundtruth A: Three matches from $1^{st}$ division and two matches from $2^{nd}$ division.
- Groundtruth B: Three matches from $2^{nd}$ division and three matches from $3^{rd}$ division.
- Groundtruth C: One match from $1^{st}$ division, one match from $2^{nd}$ division and one match from $3^{rd}$ division.

These video sequences have been produced by the Catalan television (TVC) and they belong to the Spanish soccer league. Each match lasts approximately two hours and is in MPEG-4 Simple Profile format, has a resolution of 360x288 and a GOP structure IPPPPPPP. All the results presented in the sequel have been obtained using these video groundtruths.

## 3. AUDIO AND VIDEO DESCRIPTION EXTRACTION

This section explains the techniques used to analyze the video sequences. Each approach is introduced briefly and the interested reader is addressed to the references for further details.

### *3.1 Shot Boundary Detection*

In order to generate a video summary, the original video sequence is usually split into a minor temporal unit. Depending on the type of summary this unit can range from a video scene up to a single frame [2], [8], [11]. In this paper, the proposed basic unit is the video shot, and low, mid and high level audio-visual descriptors will be used to annotate it.

Many techniques for shot boundary detection have been presented [2], [20], [21], [22] and even specific approaches have been proposed for soccer games applications [15], [17], [18], [19]. The main difficulties encountered in all approaches are the high color resemblance of soccer shots, the randomness of their motion statistics and the mixture of abrupt and gradual transitions.

In soccer live broadcasting, hard-cuts are used during the game while cross-dissolves may be used before or after the game or during the half time. In this context, the approach presented in this paper consists in employing two types of shot boundary detectors, one for hard cuts and the other for cross dissolves. The abrupt shot detector is employed during the action of the game and the dissolve detector, along with a whistle detector algorithm, in the beginning and final parts of the game. The chosen cross-dissolve boundary detector has been proposed by W. Abd-Almageed in [23] and is based on the rank analysis of a matrix composed by n-histogram frames though a Singular Value Decomposition (SVD) factorization. On the other side the hard-cut shot boundary detector consists on a traditional histogram frame-by-frame comparison through the Chi-Square distance. These shot boundary detection algorithms have been applied to the grountruth soccer video sequences explained above. In particular, for the groundtruth C a recall of 95.2% and a precision of 98.8% for the abrupt transition detector, and a recall of 91.5% and a precision of 84.4% for the cross-dissolve detector have been obtained. These results show that the selected algorithms are adequate to be used in the overall scheme.

### *3.9 Keyframe Extraction*

Once the video sequence is divided in shots, the keyframe of the shot is calculated. Keyframe extraction within a shot [7], [24] is an essential task in a summarization scheme in order to discriminate among redundant frames. In our case the goal is to select the keyframes that will be further analyzed by the color and the person description tools of the analysis bank.

The proposed approach is a novel iterative algorithm that is based on motion activity and color statistics. Initially the algorithm processes all the motion compensation vectors within a shot and performs a full search analysis in order to find first the frame with the maximum motion activity, $M_{max}$. Then, $M_{max}$ is compared against an adaptive threshold $T_1 = \alpha M_{median}$, defined as the median motion activity in the shot multiplied by a constant $\alpha$ bigger than 1. The constant $\alpha$ specifies how much deviation from the median can be accepted as a maximum. If $T_1 < M_{max}$ the shot is split into two sub-shots in the $M_{max}$ time instant and the algorithm is applied again in both sub-shots. Otherwise, if $T_1 > M_{max}$, a search analysis is performed to find the minimum motion activity, $M_{min}$. Once $M_{min}$ is found, the closest intra-frame image from the MPEG video sequence stream in that time instant is extracted and labeled as a candidate keyframe. After finding a maximum of 10 candidate keyframes the iterative algorithm stops. Finally, the color resemblance among the candidate keyframes is assessed. The keyframes that are similar in terms of color are discarded by doing a histogram comparison in the HSV colorspace using the Chi-Squared distance.

This algorithm is able to summarize each shot in keyframes depending on the intensity of action of the scene in a computationally efficient manner. Moreover it provides low-motion activity keyframes that usually yields a better performance in the person detection phase.

### *3.9 Low-level Descriptors*

Once the video sequence has been divided into shots and the keyframes have been selected, low-level descriptors are extracted to describe the audio-visual soccer content. Audio descriptors [25] play a crucial role in the summarization process of soccer sequences because the most important events tend to happen where the audio presents concrete features. For example in a goal occasion the audio power suddenly increases and in a referee's whistle instance specific harmonics are spotted at concrete frequency ranges. Motion descriptors [26] describe the level of action in the scene (e.g. the peace in a panoramic view or the high intensity movement in a goal occasion) and at the same time they can provide information about camera operations such as pannings, tiltings, zooms-in, etc. These descriptors are efficiently computed exploiting the motion compensation information provided in the MPEG streams. Color descriptors [26] in soccer events help in recognizing the type of view of the field (close-up shot, medium shot, long shot, the stands, etc.). The color descriptors have been calculated using the MPEG-7 Low Level Feature Extraction Library [40].

In order to extract low-level descriptors, and taking into account the soccer field, the following MPEG-7 descriptors have been selected:

Audio domain: The Audio WaveForm, the Audio Power, the Audio Fundamental Frequency, the Audio Spectrum Envelope and the Audio Spectrum Centroid.

Motion domain: The Motion Activity Descriptor and the Camera Motion Descriptor.

Color domain: The Color Structure Descriptor, the Dominant Color Descriptor and the Color Layout Descriptor.

### 3.9 Persons Descriptor

The detection of soccer players in close-up shots is a very important feature of the highlights summary generator system. However, in order to develop an efficient system of soccer player's detection, some constraints derived from the soccer scenarios need to be taken into account. In particular, images with low resolution and/or high luminance variation, faces with different scale factors, rotations, poses and/or occlusions, profile faces tilted up to 90º as well as high angles face shots can be found in the soccer scenario. Although results on sports players detection have been reported [27], [28], and [29], and there are a number of general face detection schemes [30] and [30], there is a need to develop more robust schemes taking into account these soccer-based constraints. In this context a robust persons descriptor based on a face detection approach has been implemented targeted to detect players in a soccer video sequence. Our approach is based on the well-known Viola and Jones AdaBoost [29] approach face detector where a skin filter has been included to decrease the number of false alarms. Let us note, that the person descriptor can also be applied to detect events where persons are involved. Fig. 2 shows some examples of close-ups detected with the proposed approach.

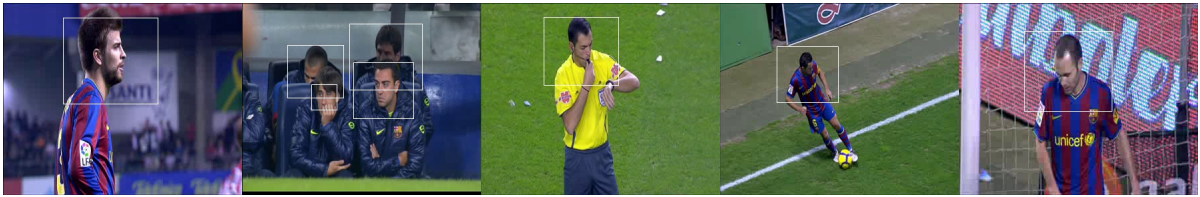

**Fig. 2** Face detections in soccer. Profile case (a). Detections in the bench and public examples (b).. Different poses and high angle shots (c), (d) and (e)

The general overview of the system is shown in Fig. 3., where several classifiers are used for the face detection stage using the approaches presented by Viola and Jones [29] and implemented in [34].

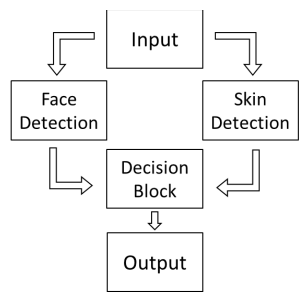

**Fig. 3** General overview of the persons descriptor system

A specific trained classifier to detect head and shoulders has also been used [35], to increase the detection rate. The skin filter is derived from the method proposed in [32] but using the RGB and HSV color subspaces instead.

Subsequently a simple condition adapted from [33] is defined for the HSV subspace to refine the detection. Finally, a decision block to reduce the false alarm level is applied. The proposed approach takes into account all the possible ethnic groups.

The recall and precision of the proposed face detection approach using groundtruth B present values of 92.8% and 89.45% respectively. In this case, to check the robustness of the approach, the test has been performed with a database formed by 550 images with multiple faces from the second and third Spanish soccer division, which present a worse scenario than first division since the TV production for these categories have a different production process.

### *3.9 Replay Detector*

The most interesting events of interest in a soccer game are usually presented through replays from different scene viewpoints or in slow motion. Hence, replay detection approaches can offer reliable descriptors in the highlights summarization process.

The TV production style in soccer media usually identifies the beginning and ending of a replay using two identical logos. Therefore, the strategy to detect a video sequence soccer replay can be based on logo detection and is generally implemented in four stages [36], [37], [38], [39]: searching for video frames that are candidates to contain a logo, detection of the logo pattern used in the soccer media, matching of the logo pattern along the soccer video and pairing the detected logos to identify the replays.

The proposed replay detector is presented in Fig. 4 and it is based on [39], that has been modified to improve the false detection rate.

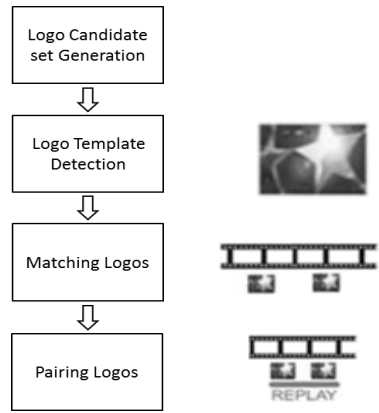

**Fig.** 4 Replay detector.

The algorithm employed in the first step uses the luminance difference between frames to detect the peaks along the video frames that characterize a logo presence. This method is suitable for hard logo transitions; however, as it presents a low performance for gradual transitions, the algorithm in [39] has been modified by aiding a shot boundary detector [23]. An algorithm, based on the k-means clustering algorithm on the luminance and variance of the frame images, is applied in order to detect the logo template.

Once the logo templates have been detected, the logos and the candidate images are converted to gray scale, and a pixel-by-pixel subtraction is performed. The sum of all the differences gives an idea about their similarity. Finally, a threshold is applied to detect the correct matches that present the lowest difference values.

The proposed approach has been tested in a database formed by different logo patterns with a total number of logos to be detected of 507 and 226 replays. The recall and precision in the logo detection process is of 100% and 99.6% respectively. In reference to the detection of the replays, the recall is 99.12% and 100% for the precision, which means some missing replays due to logos not detected. A replay is identified if only its beginning and ending logos are detected. In TV production styles that only use single logos, the pairing logos stage is not applied.

### *3.9 Zoom Detector*

The zoom operation usually indicates an important instant in a soccer match. Consequently it is essential to have a zoom detector that may help finding the highlights of the match. Several techniques for zoom detection exist in the literature [41], [42], [43]. The most common methods involve computing the optical flow, fitting it into a parametrical motion model and assessing the corresponding coefficients of the model. Unfortunately the majority of these steps include iterative techniques that require high computational costs. In this paper the proposed algorithm consists on exploiting the motion compensation vectors directly extracted from the MPEG stream and using the zoom detection method proposed by in [42]. Initially, a 23x18 motion vector field is extracted from a 360x288 MPEG video sequence and then motion vector outliers are removed using a 3x3 median filter. Afterwards the algorithm divides the motion field into the following golden section hypothesis [42]. This hypothesis splits the image into 3:5:3 proportions vertically and horizontally as shown in Fig. 5.

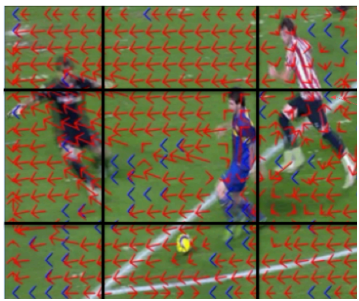

**Fig. 5** Golden sections hypothesis

Once the motion field is separated into the golden sections, the motion vector that represents the dominant direction of each region is calculated by averaging the motion vector coordinates within each area. Finally each one of the dominant vectors is parameterized into a line and if there is a common focal point among the intersections of these lines, a zoom operation is detected. This zoom detector has been tested in the groundtruth A containing 120 soccer zooms and the results are a recall of 78.33% and a precision of 88.68%. Given that in our application a higher precision rate than a higher recall is preferred, the achieved results are considered satisfactory.

### *3.9 Long Shot Detector*

In order to summarize a soccer match, the most significant highlights need to be identified. One way to simplify this process is to discard first the non-relevant occurrences. A shot that is usually of low interest in a soccer game is the one that offers a panoramic view of the field. This type of shot is denoted as *long-shot* (see Fig. 6 a,b) and is characterized by having the green as the dominant color. However, as the predominance of green color can be found also in a mid-shot, defined as in intermediate shot between a *close-up* shot and a *long shot* (see Fig. 6 c, d), specific techniques need to be developed to detect long-shots.

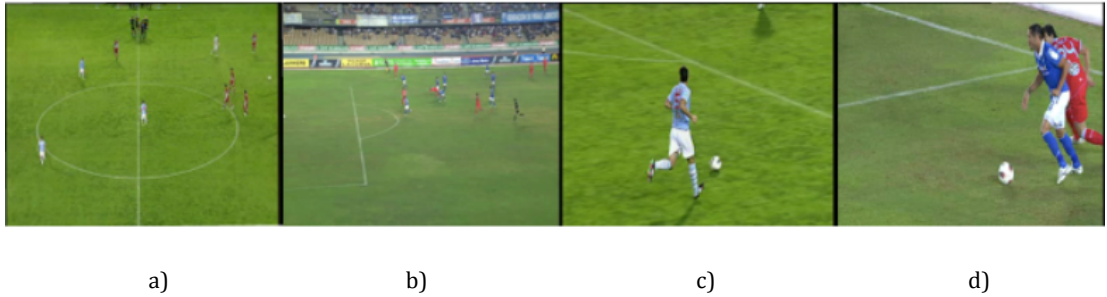

**Fig. 6** (a), (b) long-shots; (c), (d) mid-shots

The most common approaches in the literature to distinguish long-shots use as a discriminant feature the grass percentage in the whole image [44] or the grass percentage of different image regions [17], [18]. A novel approach is presented in this paper that takes into account the grass predominance in the spatial domain but also the color homogeneity in the frequency domain. The approach consists in analyzing the keyframes obtained previously through three stages: first the non-relevant information is removed, then the green dominance is assessed and finally the color homogeneity is evaluated. The initial stage consists in eliminating the majority of the non-grass pixels from the image. To do so, one-third from the top of each image is cropped, as this part of the image is not discriminant for the analysis and usually corresponds to the public.

The next stage filters the shots with green color dominance. This is done by setting rule-based thresholds on the RGB dominant colors of the image and their corresponding percentages. This information is directly extracted from the MPEG-7 Dominant Color Descriptor [26]. The third stage assesses the color homogeneity by analyzing the parameters from the MPEG-7 Color Layout Descriptor [26]. This color descriptor reduces the input image into a tiny 8x8 YCrCb image and finds the Discrete Cosine Transform (DCT) coefficients. In the *long-shot* view, soccer players are represented with few pixels; therefore the players can be removed. This procedure provides a uniform green image in the *long-shot* scenario in contrast to the *mid-shot* case. Consequently the system analyzes the homogeneity in the green color by calculating the variance of the first 9 alternate coefficients of DCT chromas and comparing them with a low threshold. The keyframes' variances that are below the threshold are marked as belonging to a *long-shot* view. The results from a total of 373 keyframes coming from the groundtruth A achieve a recall of 80.20% and a precision of 96.25%.

### *3.9 Whistle Detector*

In many sports, such as soccer, the detection of the referee's whistle provides highly valuable information to detect events of interest. Therefore, reliable and robust whistle detection is a key objective in the design of methodologies for automatic sport highlighting. Whistle detection has been considered extensively in the literature and different strategies have been proposed [45], [46], [47], [48] and [49]. Most of these proposals are based on analyzing the spectral content of the signal and detecting the maximum energy in the whistle's frequency band. Although these methods produce acceptable results in some scenarios, the number of false detections may be significant in recordings with too much cheering noise nearby the whistle's frequency band. The method proposed in this paper tries to improve these results by further exploiting the spectral characteristics of professional whistles.

The proposed strategy for whistle detection is presented in Fig. 7.

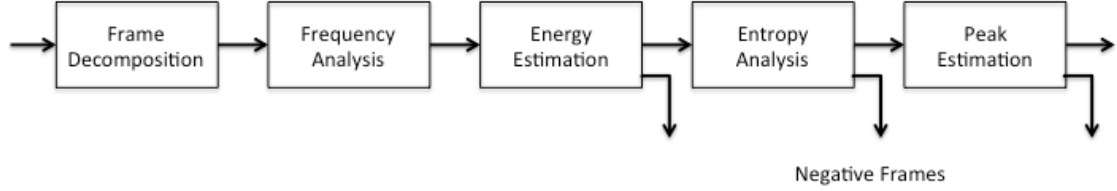

**Fig. 7** Whistle detector algorithm

The first block segments the audio signal onto frames. In our database, a sampling frequency of 48 kHz, and a frame size of 4,800 samples have provided accurate results for the next stages.

The frequency analysis block computes a set of Discrete Fourier Transform (DFT) samples in the band of interest for every audio frame. That band is selected between 3.5 kHz and 4.5 kHz as broadly include the frequencies produced by professional whistles. The reduced set of samples of the DFT is computed using the Goertzel algorithm [50].

The next stage estimates the energy contained in the interest band of an audio frame. Once this energy is estimated, a threshold is applied to detect the whistle. Although there exists high correlation between energy peaks and whistles, in some difficult scenarios thresholding produce an unacceptable number of false alarms due to the presence of noise in the band of interest. Then, the last two stages of the whistle detector system try to reduce the number of false detections by exploding the fact that the whistle spectrum is made of 3 tones at very close frequencies. However, as these frequencies are not stable and may vary slightly with the specific excitation, the direction of blowing and the whistle model, in order to discriminate the tonal vs non-tonal nature of the audio frame, an entropy-based stage is introduced.

The approach consists in considering a normalized version of spectrum samples in the interest band as being samples of a probability density function. When the entropy is computed on these samples a number that indicates the spread of the samples in the frequency band will be obtained. High numbers indicate wide spread while small numbers indicate that the energy is concentrated on a few, high probable energy samples. This entropy-based concept is defined as:

$$H(m) = \sum_{k=K1}^{K2} \rho_m(k) \log_2 \rho_m(k) \quad \text{with } \rho_m(k) = \frac{|X_m(k)|^2}{\sum_{r=K1}^{K2} |X_m(r)|^2} \tag{1}$$

where K1 and K2 represent the DFT samples at the limits of the interest region and m represents the frame time index. Applying a threshold on this entropy value permits to discard all those audio frames where the energy is spread over the interest band. The tonal audio frames that do not exceed that threshold will be further processed by the final stage. Entropy analysis rejects some of the false positives in the example. Finally the last stage consists in discarding some sounds that are sometimes confused with the whistle by selecting the total number of peaks that exceed a threshold. Only audio frames with 2 or 3 peaks are finally validated.

In order to test the performance of the whistle detector a database has been created by manually annotating the complete recordings from groundtruth A. The annotation is performed using both video and audio tracks. Video track gives a helpful context required to discern referee's whistles among other sounds such as vuvuzelas, horns, supporter's whistles, etc. Aside from this database a test signal has also been generated. This test signal is made up of a selection of referee's whistles annotated in the database in conjunction with other especially difficult sounds, usually inducing false positives, which have been encountered in these recordings. The total length of the test signal is around 60 s and contains 10 referee's whistles, which are represented in Fig. 8a. Fig. 8b represents the energy estimate. The result of applying the entropy function to the test signal is represented in Fig. 8c where it can be verified that the entropy decreases in the frames that correspond to whistles. The results obtained when applying the algorithm to the complete database show a recall of 93% and a precision of 88%.

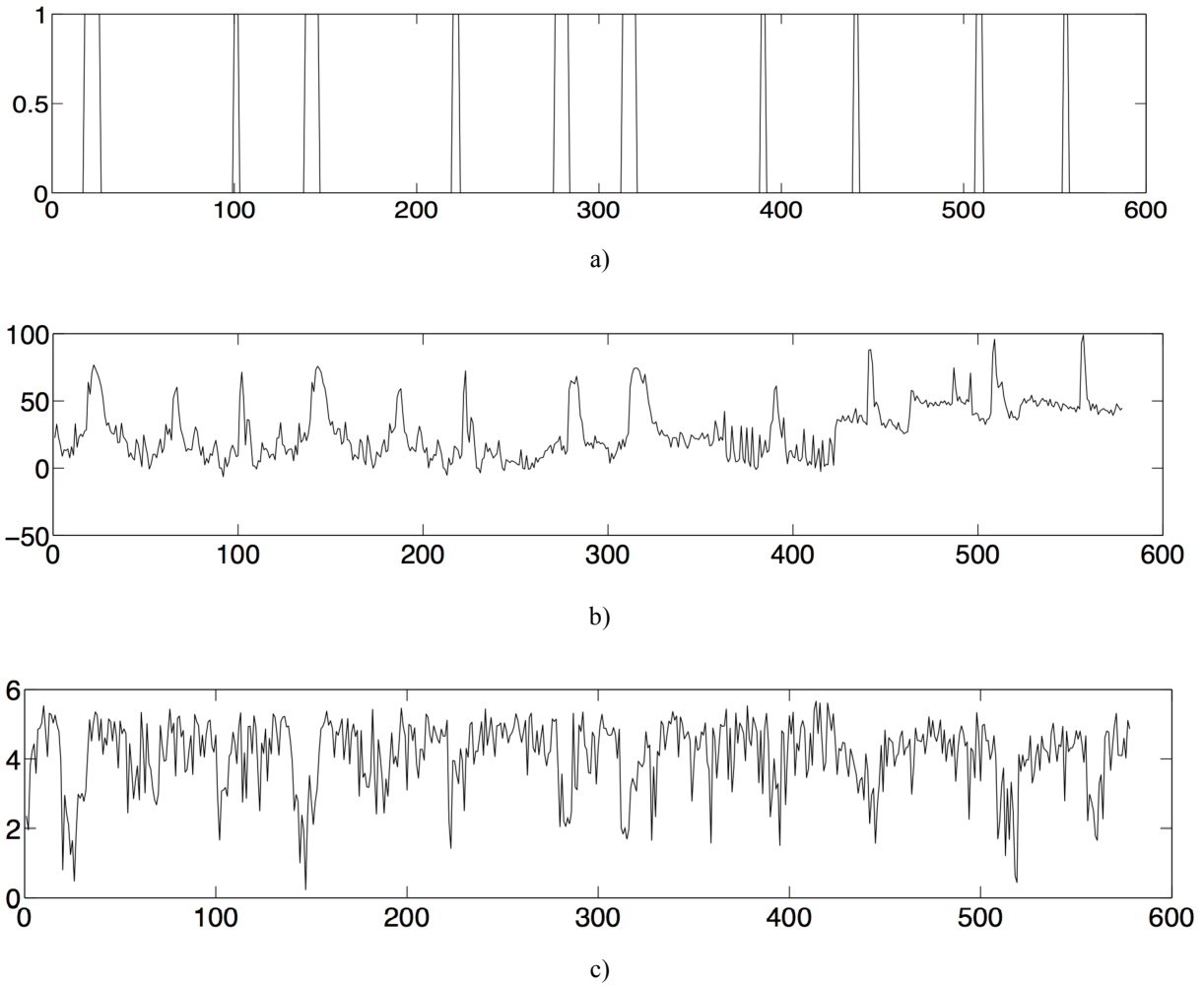

**Fig. 8** Summary of results obtained by the different modules of the whistle detector for the test signal.
a) Groundtruth b) Energy in the interest band c) Entropy estimation of the spectrum

### 3.9 Inter and Intra shot-based audio detectors

Temporal audio changes play a crucial role in the detection of relevant soccer events. Hence, to spot important instants in the evolution of the audio, three detectors that measure audio power variations have been extracted:

• The A.Power.H and A.Power.VH detectors represent peak levels of audio power within a shot, where H stands for high and VH for very high. Logical binary values are associated to those shots whose maximum audio power is over 95% and 97%, respectively, in reference to the maximum audio power value of the entire audio soccer track.
• A.IntraInc.50 and A.IntraInc.100 represent audio power increments within a shot. The low-level audio power descriptors are averaged for every second. The first detector is used to represent logical true values when the audio power in these averaged intervals is increased in 50% whereas the second detector represents increments of 100%.
• A.InterInc.50 and A.InterInc.100 are the same as A.IntraInc.50 and A.IntraInc.100 but they refer to average audio power increments between contiguous shots.

These basic low-level descriptors will provide helpful and insight information to generate the video soccer match summarization when combined with other descriptors.

# 4. HIGHLIGHT GENERATOR ARCHITECTURE

The goal of this section is to define the tool that provides the video sequence summary of soccer highlights. The inputs of this tool are the low level audio-visual descriptors introduced in Section 3. The highlight generator is shown in Fig. 9 and consists of a filter bank and an event detector. First the filter bank scores each shot using a set of elementary and advanced filters, then the event detector generates the final summary by assessing the previous scores and a set of summary options defined by the user.

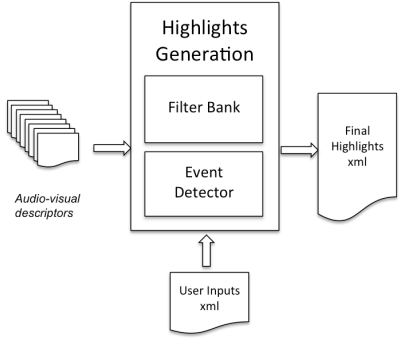

**Fig. 9** Highlight generator architecture diagram

The filter bank is composed of elementary and advanced filters. Elementary filters are defined as filters that detect or classify events in each shot by using features directly provided by the low-level audio-visual descriptors. Filters that fall into this category are: the long shot detector, the zoom detector, the whistle detector, the replay detector, the persons detector, the high motion detector, the audio power detector, the audio intra power shot detector, the audio inter power shot detector, the long duration classifier, the medium duration classifier, the short duration classifier and the very short duration classifier. Duration classifiers categorize the length of a shot. The rest of the detectors have already been detailed in Section III.

The objective of the advanced filters is to provide a higher-level description of a shot by linearly combining elementary filters. To do so, each elementary filter is understood as a boolean function $F$, which is weighted by a coefficient $\omega$. These $\omega$ coefficients can take positive or negative values in order to benefit or penalize the outputs of elementary filters. The result of an advanced filter operation is a *local score* $L$ and their aggregation gives the *global score* $G$ of that shot as defined in (1),

$$G_s = \sum_{j=1}^{M} L_j = \sum_{j=1}^{M} \sum_{i=1}^{N} \omega_{i,j} F_i \tag{2}$$

where *M* and *N* are the total number of advanced and elementary filters respectively.

Advanced filters can be interpreted as mid-level event detectors that may be used to identify goals, faults, important moments, goal-celebrations, etc. In our results, the weights of the elementary filters are selected manually based on previous knowledge about production and edition techniques. Alternatively, the weights can be acquired by training a linear classifier.

Once all the shots are scored, the event detector stage is initiated. Here the aim is to choose the final shots that will produce the video sequence of soccer highlights. In order to do so this stage takes into account the user's preferences about the summary. The user can specify in advance the length of the summary and also the percentage of appearance of each one of the advanced filters. As an example, the user could select a summary of 3 minutes where 50% of the time is dedicated to goals, 30% to important game moments, 10% to goal-celebrations and 10% to faults. The shot selection algorithm gathers the shots with the highest local scores for each one of the advanced filters, sorts them by their global scores and finally selects a number of shots till the desired duration is achieved. One problem that commonly arises is that there are too few shots for a specific advanced filter. In this case the system equally distributes the missing duration within the remaining advanced filters.

The output of the summary generator is a XML file that contains the time codes of each selected shot.

## 5. HIGHLIGHT GENERATOR PERFORMANCE

In this section classified shot distributions of automatically generated summaries are shown and an example of an advanced filter, a goal-scoring detector filter, is defined and its performance assessed. The database used to evaluate the performance of the system is the groundtruth A described in Section II. Similar results have been found for other groundtruth sequences.

In order to assess a summary, all the resulting shots in it have been classified in five groups: goals, offense/defense, offside-faults, persons and replays. The goals category contains the shots where a goal or a goal opportunity happens; opportunity of goal is any offensive play that ends close to the goal or due to a goalkeeper intervention. Offense/defense represents the offensive and defensive actions that do not entail a goal or goal opportunity. Offside/faults groups faults and offside. Persons represent the shots that contain public or persons as soccer players, coaches and other team members. Replays are shots that contains replays of faults, offside, goals and other interesting events.

Aiming to imitate manually annotated Catalan soccer summaries of 15 minutes, a specific percentage of appearance for each advanced filter in the event detector stage has been devised. The resulting distribution of classified shots of two of these soccer summaries is presented in Fig. 10.

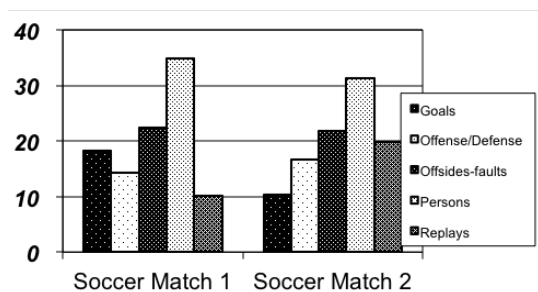

**Fig. 10** Classified shot distribution of two soccer matches

It can be observed that the highlight generator is quite flexible, since the shot distribution can be easily modeled depending on the match according to a specific production style. The person's category is the most popular shot type, in contrast to goals and replays that rely on the soccer match activity. Moreover offense/defense plays an important role in long soccer summaries, as this is where most of the action takes place.

An example of an advanced filter is a goal-scoring detector. Goal scoring is one of the most interesting and relevant actions in soccer. However, its high semantic level makes its identification very difficult if only low-level descriptors are available. In order to overcome this, a statistic study has been conducted and a pattern has been devised combining low-level and mid-level descriptors. A goal is defined as an abrupt increment of the audio power (A.Power and A.Intra) followed in the near shots by the soccer player's celebration (Persons) and/or a replay detection (Replay). Table 1 contains the results of the detected goals for the soccer matches presented in Section 2.

**Table 1** The goal filter performance.

| Goals | Match 1 | Match 2 | Match 3 | Match 4 | Match 5 |
|---|---|---|---|---|---|
| Total Number of Goals | 2 | 4 | 3 | 6 | 3 |
| Total Detected Goals | 1 | 4 | 3 | 4 | 1 |
| Goals Not Detected | 1 | 0 | 0 | 2 | 2 |

The proposed system is able to detect satisfactorily over the 70% of the total amount of goals of the five soccer matches analyzed. However its performance presents some unpredictable limitations due to changes in TV production style or in external uncontrolled factors as for instance the public that influences the audio power producing peaks and/or significant increments. It is important to highlight that soccer match 4 and 5 present a higher number of goals not detected and the main reason stems from the fact that these sequences do not contain replays.

## 6. CONCLUSIONS

This paper presents a novel framework for generating automatic video highlights of soccer broadcasting video sequences. The proposed approach consists on segmenting the video sequence into shots and scoring them using a set of low-level and mid-level audio-visual descriptors. Of special interest in the approach is the use of the audio information that provides additional robustness to the overall performance of the summarization system. Low-level descriptors describe audio, color and motion through MPEG-7 standards and mid-level descriptors annotate persons, whistles, zooms, long-shots and replays. The scoring process linearly combines and weights each descriptor by taking into account end-user preferences and custom soccer highlights are then generated. The proposed framework has been tested in a specific scenario and satisfactory results have been achieved.